\documentclass[11pt,twoside]{article}
\usepackage{asp2004}
\usepackage{psfig}
\usepackage{epsf}
\usepackage{graphics}
\usepackage{lscape}
\begin{document}

\newbox\grsign \setbox\grsign=\hbox{$>$} \newdimen\grdimen \grdimen=\ht\grsign
\newbox\simlessbox \newbox\simgreatbox
\setbox\simgreatbox=\hbox{\raise.5ex\hbox{$>$}\llap
     {\lower.5ex\hbox{$\sim$}}}\ht1=\grdimen\dp1=0pt
\setbox\simlessbox=\hbox{\raise.5ex\hbox{$<$}\llap
     {\lower.5ex\hbox{$\sim$}}}\ht2=\grdimen\dp2=0pt
\def\simgreat{\mathrel{\copy\simgreatbox}}
\def\simless{\mathrel{\copy\simlessbox}}

\title{Theory of Outflows in Cataclysmic Variables}
\author{D. Proga}
\affil{JILA, University of Colorado, Boulder CO 80309, USA\\
Present address: Department of Astrophysical Sciences, Princeton
University, Peyton Hall, Princeton NJ 08544, USA }

\begin{abstract}
We review the main results from line-driven (LD) models of winds
from accretion disks in cataclysmic variables (CVs). We consider
LD disk wind models in the hydrodynamic (HD) and
magnetohydrodynamic (MHD limits. We discuss the basic physical
conditions needed for a disk wind to exist and the conditions for
the wind to be steady or unsteady. We also discuss how the
line-driven (LD) wind structures revealed in numerical simulations
relate to observations. In particular, we present synthetic line
profiles predicted by the LD wind models and compare them with
observations. Our main conclusion is that, despite some problems,
line-driving alone is the most plausible mechanism for driving the
CV winds. This conclusion is related to two facts:  1) LD wind
models are likely the best studied wind models, analytically and
numerically, and 2) it is most likely that the predictive power of
the LD wind models is much higher than of any other wind model so
far. Preliminary results from LD-MHD wind models confirm that
magnetic driving is likely an important element of the wind
dynamics. However, magnetic driving does not seem to be necessary
to produce a CV wind. The most important issues which need to be
addressed by future dynamical models, regardless of driving
mechanism, are the effects of the position-dependent
photoionization and the dynamical effects in three dimensions.
\end{abstract}

\section{Introduction}

Winds in cataclysmic variables (CVs) exemplify very well a
phenomenon of accretion disks being accompanied by mass outflows.
Other systems where this phenomenon occurs are active galactic
nuclei (AGN) and young stellar objects (YSOs). In the case of CVs,
key evidence for outflows comes from P-Cygni profiles of strong UV
lines such as C~IV$\lambda$1549. However, the evidence for the
winds is not limited just to the strong UV lines \cite{LK}.
Understanding the winds in CVs is important on its own right
and because they have been the best observed outflows from compact
objects and promise to provide us with insights into all disk
outflows. The interpretation of data is usually based on fitting
observed profiles to synthetic profiles calculated from kinematic
models \citep[e.g.,][]{MR87,Drew,SV,KWD95,LK}. We refer a reader
to Froning (2004, this volume) for a review of the observations
and interpretations of CV winds.

Magnetic fields, the radiation force and thermal expansion have
been suggested as mechanisms that can drive disk winds. These
three mechanisms have been studied extensively using analytic as
well as  numerical methods. As a result of these studies,
theoretical models have been developed that allow us to estimate
under what physical conditions each of these mechanisms is
efficient in launching, accelerating and collimating disk
outflows. Here we will focus on models of radiation-driven disk
winds (section 2) and a hybrid model in which both radiation and
magnetic driving is considered (section 3).

\section{LD HD Models}

More than three decades of studies of winds in hot stars provide
us with a very good understanding of how line-driving  produces
powerful high velocity winds (e.g., \citeauthor{CAK}
\citeyear{CAK}, hereafter CAK; \citeauthor{FA} \citeyear{FA};
\citeauthor{PPK}\citeyear{PPK}). The key element of the
\citeauthor*{CAK} model is that the momentum is extracted most
efficiently from the radiation field via line opacity. The
Eddington limit, $L_{\rm Edd} = 4\pi cGM/\sigma_{\rm e}$, is the
maximum luminosity a spherical object of mass $M$ may achieve
before the radiation pressure mediated by photons scattering off
free electrons becomes so large as to drive off the object's
atmosphere and envelope.  It is commonly the case that the
effective cross section for photon scattering is greatly increased
by the presence of Doppler-shifting bound-bound transitions.
\citeauthor*{CAK} showed that the radiation force due to lines,
$F^{rad,l}$ can be stronger than the radiation force due to
electron-scattering, $F^{rad,e}$ by up to several orders of
magnitude (i.e., $F^{rad,l}/F^{rad,e} < M_{\rm max} \approx
2000$). Thus even a star that radiates at around 0.05\% (i.e.,
$1/M_{\rm max}$) of its Eddington limit can have a strong wind.

Early models of radiation-driven disk winds have applied
assumptions that either restrict the flow geometry or require the
flow to be time-independent [e.g. \citet{VS} and a model for AGN
wind by \citet{MCGV}]. Recently numerical models of 2.5-D,
time-dependent radiation driven disk winds have been constructed
for application to, in the first instance, CV disk winds (e.g.,
\citeauthor{PKB97}\citeyear{PKB97}; \citeauthor{PSD98}
\citeyear{PSD98}, hereafter PSD~98; \citeauthor{PSD99}
\citeyear{PSD99}, hereafter PSD~99;
\citeauthor{P99}\citeyear{P99}). In particular,
\citeauthor*{PSD98} adopted numerical techniques to integrate the
coupled HD and radiation transfer equations in order to study the
multidimensional and time-dependent character of line-driven (LD)
disk winds from first principles. For the HD equations, they used
the well-tested ZEUS code \citep{SN92} extended by the addition of
a term in the equation of motion which accounts for the radiation
force due to spectral lines of the form
\begin{equation}
    F^{rad,l} = \int_{\Sigma} \left(\frac{\sigma_{\rm e} d{\cal F}}{c}\right)
         M(t) .
\end{equation}
The term in brackets is the electron scattering radiation force
($F^{rad,e}$) and $M$, the force multiplier, is the increase in
opacity due to spectral lines. The integration is over all visible
radiating surfaces ($\Sigma$). Note that $d{\cal F}$ contains the
total frequency-integrated intensity emitted at a given location.
\citeauthor*{PSD98} adopted the simple form for $M$ which still
underpins much modeling of OB star winds, i.e. $M = kt^{-\alpha}$,
where $t$ is proportional to the local density divided by the
local velocity gradient, and $k$ and $\alpha$ are constants
(\citeauthor*{CAK}). The maximum values of $M$, $M_{\rm max}$
determines the actual luminosity limit (the effective Eddington
limit, $L_{\rm E}/M_{\rm max}$) for which the radiation pressure
mediated by photons becomes large enough  to drive off the
object's atmosphere and envelope. Note that integration over angle
of a nonlinear function of the velocity gradient tensor and the
radiation flux is required to evaluate equation (1);
\citeauthor{PSD98} took great care in the numerical evaluation of
these integrals by adopting angle-adaptive quadrature methods.
\citeauthor*{PSD98}'s formalism allows the radiation from the
central accreting star to be included both as a direct contributor
to the radiation force and as an indirect component via disk
irradiation and reemission.  To spatially resolve the flow, we
used a non-uniform (up to 200 $\times$ 200) grid in which the
subsonic acceleration zone near the disk or stellar surface is
well sampled.

\begin{figure}[!ht]
\plotfiddle{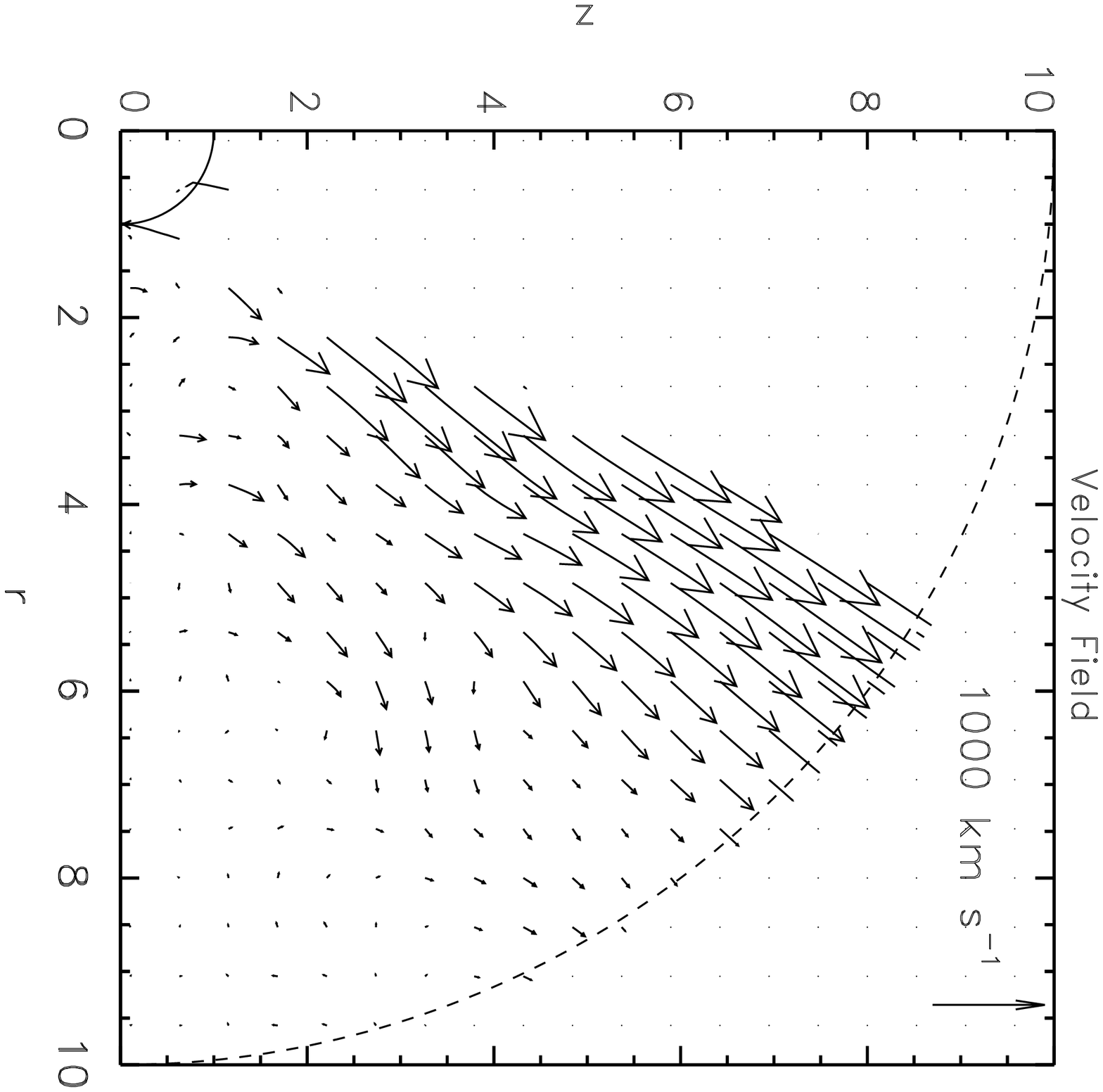}{9cm}{90}{32}{32}{50}{90}
\plotfiddle{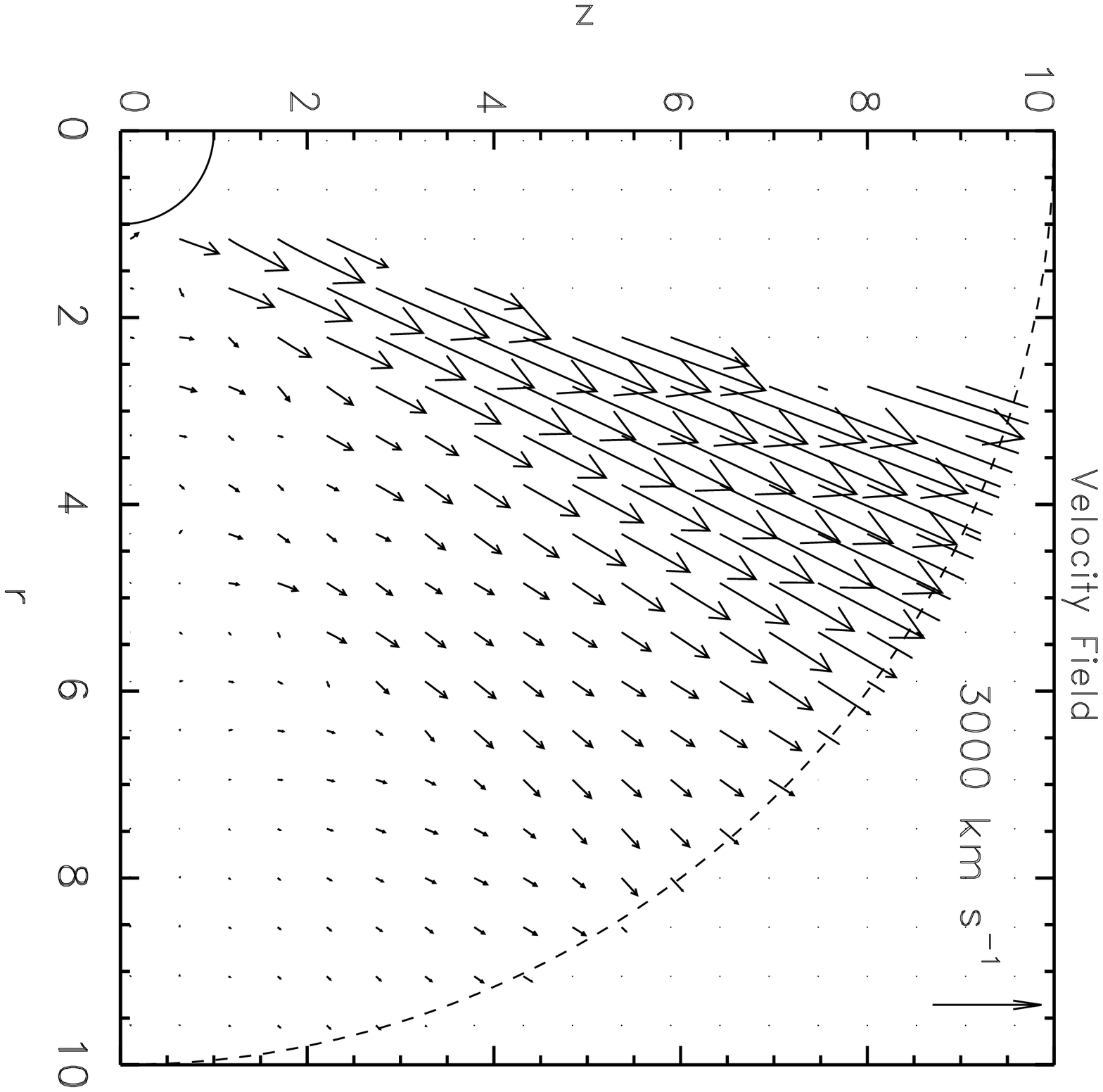}{0cm}{90}{32}{32}{230}{115}
\plotfiddle{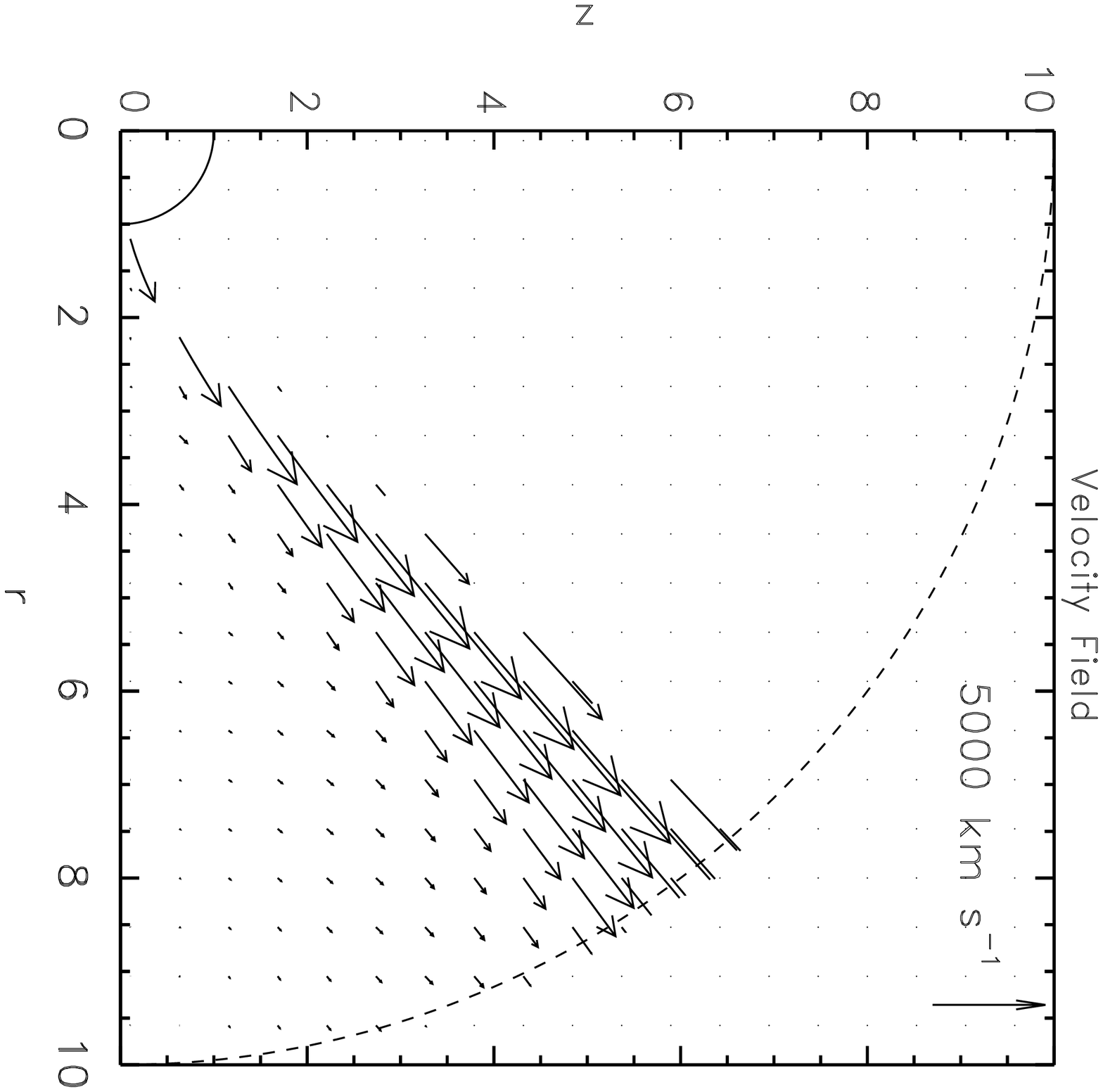}{0cm}{90}{32}{32}{50}{-40}
\plotfiddle{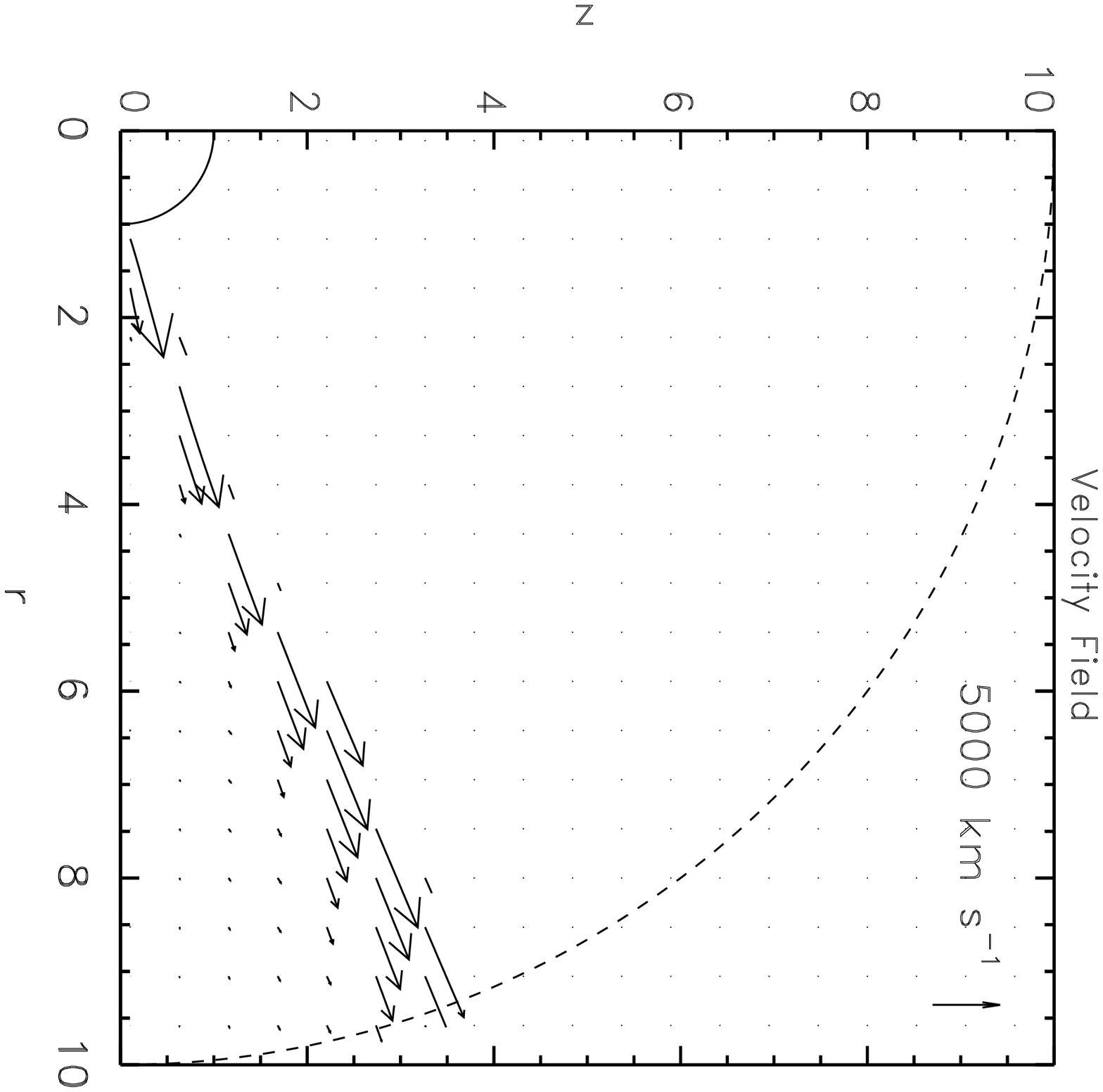}{0cm}{90}{32}{32}{230}{-15} \caption{
Maps of poloidal velocity for a range of LD disk wind models
(Figure 2 in \citeauthor*{PSD99}). The top panels are both models
with $x=0$ but with $\dot{M}_{\rm a} =10^{-8}~\rm
M_{\odot}\,yr^{-1}$ (the left hand side panel) and $\dot{M}_{\rm
a} =\pi \times 10^{-8}~\rm M_{\odot}\,yr^{-1}$ (the right hand
side panel). The bottom two panels are results for models both
with $\dot{M}_{\rm a} = \pi \times 10^{-8}~\rm
M_{\odot}\,yr^{-1}$, but with $x=1$ (the left hand side panel) and
$x=3$ (the right hand side panel). The top two panels show the
effect on the outflow geometry of increasing the disk luminosity
alone, while the top right and bottom two panels show the effect
of adding in an increasingly larger stellar component ($x$ $=$ 0,
1 and 3) to the radiation field.  Adding in an increasingly large
stellar component causes the outflow to become more equatorial.
Note that we suppress velocity vectors in regions of very low
density (i.e., $\rho$ less than $10^{-20}$~g~cm$^{-3}$).}
\end{figure}

The primary outcome of the numerical studies is the
confirmation that line driving can produce a
supersonic, biconical wind from an accretion disk in CVs.

\citeauthor*{PSD98} explored the impact upon the mass-loss rate,
$\dot{M}_{\rm w}$ and outflow geometry caused by varying the
system luminosity and the radiation field geometry. In their
study, the system luminosity, $L$ was defined as the sum of the
disk luminosity and the central star luminosity, $L_{\rm D}$ and
$L_\ast$, respectively (i.e., $L=L_{\rm D}+L_\ast=(1+x) L_{\rm
D}$, where $x=L_\ast/L_{\rm D}$). A striking outcome was that
winds driven from, and illuminated solely by, an accretion disk
yield complex, unsteady outflow (see the top panels in Figure 1).
In this case, time-independent quantities can be determined only
after averaging over several flow timescales. On the other hand,
if  winds are illuminated by radiation mainly from the central
object, then the disk yields steady outflow (see the bottom panels
on Figure 1). \citeauthor*{PSD98} also found  that $\dot{M}_{\rm
w}$ is a strong function of the total luminosity, while the
outflow geometry is determined by the geometry of  the radiation
field. For high system luminosities, the disk mass-loss rate
scales with the luminosity in a way similar to stellar mass loss
(e.g., compare the crosses and the solid-line curve on Figure 2).
As the system luminosity decreases below a critical value (the
Eddington factor, $\Gamma\equiv L/L_{\rm Edd}$ about twice
$1/M_{\rm max}$) the mass-loss rate decreases quickly to zero.

The simulations also showed that regardless of the radiation
geometry,  the two-dimensional structure of the wind consists of a
dense, slow outflow that is bounded on the polar side by a
high-velocity  stream (respectively `slow wind' and `fast stream',
for short; see e.g., Figure 1). Matter is fed into the fast stream
from within a few central object radii. In other words, the
mass-loss rate per unit area decreases sharply with radius. The
terminal velocity of the  stream is similar to that of the
terminal velocity of a corresponding spherical stellar wind, i.e.,
$v_\infty \sim {\rm a~few}~v_{\rm esc}$, where $v_{\rm esc}$, is
the escape velocity from the photosphere. Thus the difference in
geometry changes the wind geometry and time behavior but has less
effect on $\dot{M}_{\rm w}$ and $v_{\infty}$.

\citeauthor*{PSD99} applied a more accurate treatment of the line
force than \citeauthor*{PSD98} by including all the terms in the
velocity gradient tensor. Qualitative features of the new models
are very similar to those  calculated by \citeauthor*{PSD98}. In
particular, the more accurate calculations showed that models
which displayed unsteady  behavior in \citeauthor*{PSD98} are also
unsteady with the new method, and gross properties of the winds,
such as mass-loss rate and characteristic velocity are not changed
by the more accurate approach. The largest change caused by the
new method is in the disk-wind opening angle: winds driven only by
the disk radiation are more polar with the new method while winds
driven by the disk and central object radiation are typically more
equatorial.

\subsection{Nature of Unsteady Outflow}

The fact that the unsteady behavior observed in
\citeauthor*{PSD98} models has not changed with a more accurate
treatment of the radiation force indicates it is indeed a robust
property of line driven winds from disks. Why does increasing the
radial component of the radiation force 'organize' the wind into a
steady state?

\citeauthor*{PSD99}, presented the following explanation; let $r$
and $z$ define position along a streamline in the wind in
cylindrical coordinates. An increase of the vertical component of
the gravity,
$
g_z \propto -{z}/{(r^2+z^2)^{3/2}}
$
with height at a fixed radius $r$ is the main driver of the
unsteady flow. (We add that for a disk wind driven by the disk
radiation is vertical and its density can decrease downstream only
due to an increase in the velocity as there is no geometrical
dilution.) The increase of the gravity with height can be
significantly reduced if the streamlines are directed outwards
from purely vertical (i.e., $r$ increases with $z$).  At the same
time, this tilt also brings into play an increase of the
horizontal effective gravity, $g_r$, along each streamline:
$
g_r \propto [{r_{\rm f}}/{r^3}-{r}/{(r^2+z^2)^{3/2}}],
$
where $r_{\rm f}$ is the radius on a Keplerian disk at which a
streamline originates. However the increase of $g_r$ with $r'$ is
slower than the increase of $g_z$ with $z'$ because of the
decaying centrifugal term. In other words, the line force can more
easily maintain domination over gravity if the flow climbs the
gentler gravitational hill in the horizontal direction as compared
with the vertical direction. Furthermore, driving material along
streamlines outward from the vertical causes density to decline
with radius due to geometrical dilution alone as $r^{-p}$ with
$p\approx 0.5-1.5$ --- this, very usefully, tends towards
increasing the line force, thereby facilitating a better match
with trends in gravity.

We note that \citeauthor{PSD98}'s discovery of unsteady flow in
models for $x=0$ has not been confirmed by Pereyra and his
collaborators in their numerical simulations
\citep{PKB97,PKB00,PK03}. Recently, \cite{Petal04} presented
mathematically simple models and used them to argue against the
above explanation of unsteady nature of LD disk wind for $x=0$. In
particular, Pereyra et al. claimed that a gravitational force
initially increasing along the wind streamline, which is
characteristic of disk winds, does not imply an unsteady wind.
However, we find that \citeauthor{Petal04} omitted the fact that
the line force and consequently wind streamlines are coupled to
the gravitational force and radiation field. We also note that
\cite{VS} studied analytically LD driven disk wind model where
streamlines were vertical near the disk and diverging to spherical
at large distances. \citeauthor{VS} found that for a steady state
solution to exist it is necessary to enforce a radiation force
term to increase with height. \citeauthor*{PSD98} and
\citeauthor*{PSD99}'s results are then consistent with
\citeauthor{VS}'s result because for the case where the wind is
vertical near the disk only unsteady solution exist
(\citeauthor*{PSD98} computed the radiation force
self-consistently and did not enforce any special increase of it
with height [See however \citet{PKB97}]. In summary, we conclude
\citeauthor{Petal04}'s mathematically simple models, if anything
support \citeauthor*{PSD98} and \citeauthor*{PSD99}'s
interpretation of unsteady winds: when one considers the radiation
and flow geometry appropriate for the $x=0$ case (i.e., radiation
and flow are vertical near the disk) then analytic analysis
similar to that used by \citeauthor{Petal04} shows that there is
no critical point (using the terminology from analytic models of
LD stellar winds as adopted by \citet{Petal04}). However,
significant central radiation (as in the $x >0$ cases) makes a
dynamical and geometrical change in the wind solution i.e.,
increases the wind inclination angle and a critical point can
exist.

\subsection{Testing Models Against Observations}

Applying \citeauthor*{PSD98} and \citeauthor*{PSD99}'s dynamical
models to CVs one finds that radiation driving can produce disk
winds consistent with the
following observed properties of CV winds:
(i) the flow is biconical rather than equatorial as required by the
absence of blueshifted line absorption from the spectra of eclipsing
high-state CV,
(ii) the wind terminal velocity is comparable to the escape velocity from the
surface of the white dwarf (WD), and
(iii)
the spectral signatures of mass loss show a sharp cut-off as the total
luminosity in dwarf novae declines away from maximum light through the regime
theoretically identified as likely to be critical.\\
Additionally,  LD disk wind models may explain the highly unsteady
and continuously variable nature of the supersonic outflows in the
NL binaries BZ~Cam and V603 Aql \citep{P00a,P00b}.  The presence
of a slow, dense transition region between disk photosphere and
outflow in V347~Pup and UX~UMa \citep{SVM,KD} may also be
accommodated within these same models.

Building upon these numerical models, one can compare the results
of these models with the analytic results readily derived for
spherically-symmetric winds \citep[e.g.,][]{P99}. This comparison
shows the achievable disk-wind mass loss rates differ only from
the well-known one-dimensional analytic values by a factor (of
geometric origin) of order unity. A very similar conclusion was
reached by \cite{FS} who compared their analytic LD disk wind
models the two-dimensional numerical simulations of
\citeauthor*{PSD98}) and found an overall good agreement in the
streamline shape, tilt angle, and mass-loss rate. These agreement
between numerical models and analytic ones allow us to generalize
beyond the limited set of models for which numerical results
already exist.

\begin{figure}[!ht]
\plotfiddle{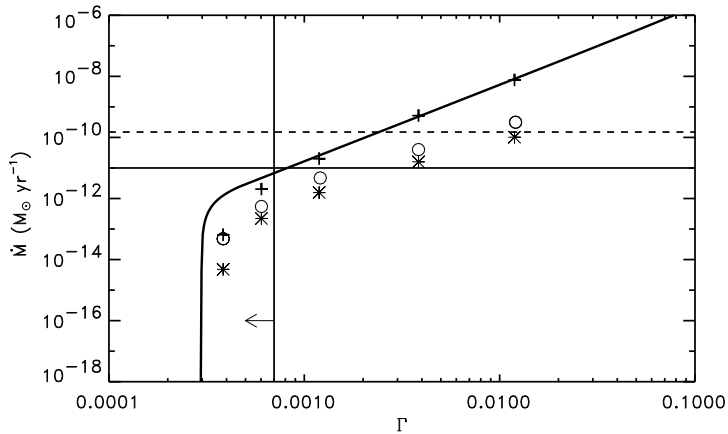}{5.0cm}{90}{46}{46}{205}{-65} \caption{
Model mass loss rates for radiation-driven disk winds as a
function of $\Gamma$ where $\Gamma \equiv L/L_{\rm Edd}$ [Figure 1
in \cite{DP00}].  The solid-line curve is the 1-dimensional mass
loss rate, derived analytically, in the case that ${M}_{{\rm max}}
= 4000$ and $\alpha = 0.4$.  Also shown are numerical disk-wind
mass loss rates for the same ${M}_{{\rm max}}$ for $\alpha = 0.4$
(crosses), $0.6$ (circles), $0.8$ (asterisks).  (See \citep{P99}
for further details).  The vertical line superimposed is drawn at
$\Gamma$ corresponding to a mass accretion rate of $10^{-8}$
M$_{\odot}$~yr$^{-1}$ onto a CO white dwarf of mass 1~M$_{\odot}$
-- presently mass accretion rates are believed to be less than
this. The solid horizontal line typifies current
$\dot{M}$~$\xi_{C^{3+}}$ estimates, while the dashed line is the
lower limiting mass loss rate construed from wind ionization
models. }
\end{figure}

For example, \cite{DP00} compared mass loss rates predicted by the
models with observational constraints (Figure 2). They concluded
that either mass accretion rates in high-state CVs are higher than
presently thought  by a factor of 2-3 or that radiation pressure
alone is not quite sufficient to drive the observed hypersonic
flows. The difficulty in accounting for the mass loss rate in a
pure LD disk wind model for CVs is simply a reflection of the fact
that the CV luminosities just barely satisfy the basic
requirement, i.e., $L_{\rm UV}\simless 7 \times 10^{-4} L_{\rm
Edd}$. A very similar conclusion has been reached by \cite{MR00}
who analyzed observations of OY Carinae taken by the Extreme
Ultraviolet Explorer. \citeauthor{MR00} argued that line driving
alone falls an order of magnitude short of driving the observed
mass-loss rate.

Generally, one can argue that in all accretion disks, with the UV
luminosity, $L_{\rm UV} \simgreat$ a few $10^{-4} L_{\rm Edd}$
mass outflows have been observed \citep{P02}. For example,
accretion disks around: massive black holes, WDs (as in AGN and
CVs with $L_{\rm UV} \ga 0.001 L_{\rm Edd}$) and low mass YSOs (as
in FU Ori stars with $L_{\rm UV} \ga$ a few $\times~ 0.01 L_{\rm
Edd}$) show powerful fast winds. Systems that have too low UV
luminosities to drive a wind include accretion disks around
neutron stars and low mass black holes as in low mass X-ray
binaries and galactic black holes. These systems indeed do not
show outflows similar to those observed in CVs, AGN and FU Ori.

\begin{figure}[!ht]
\plotfiddle{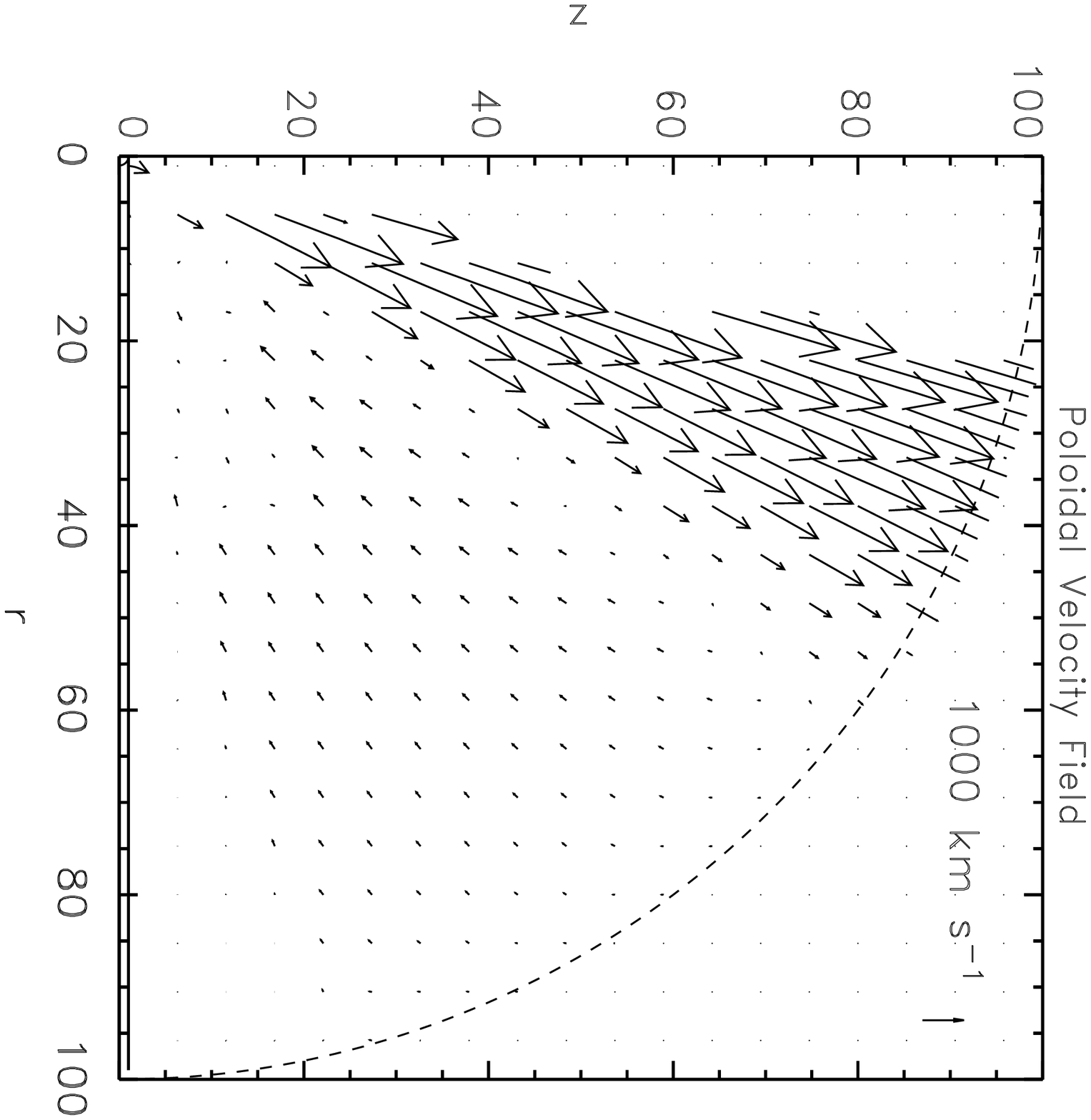}{8.2cm}{90}{28}{28}{30}{90}
\plotfiddle{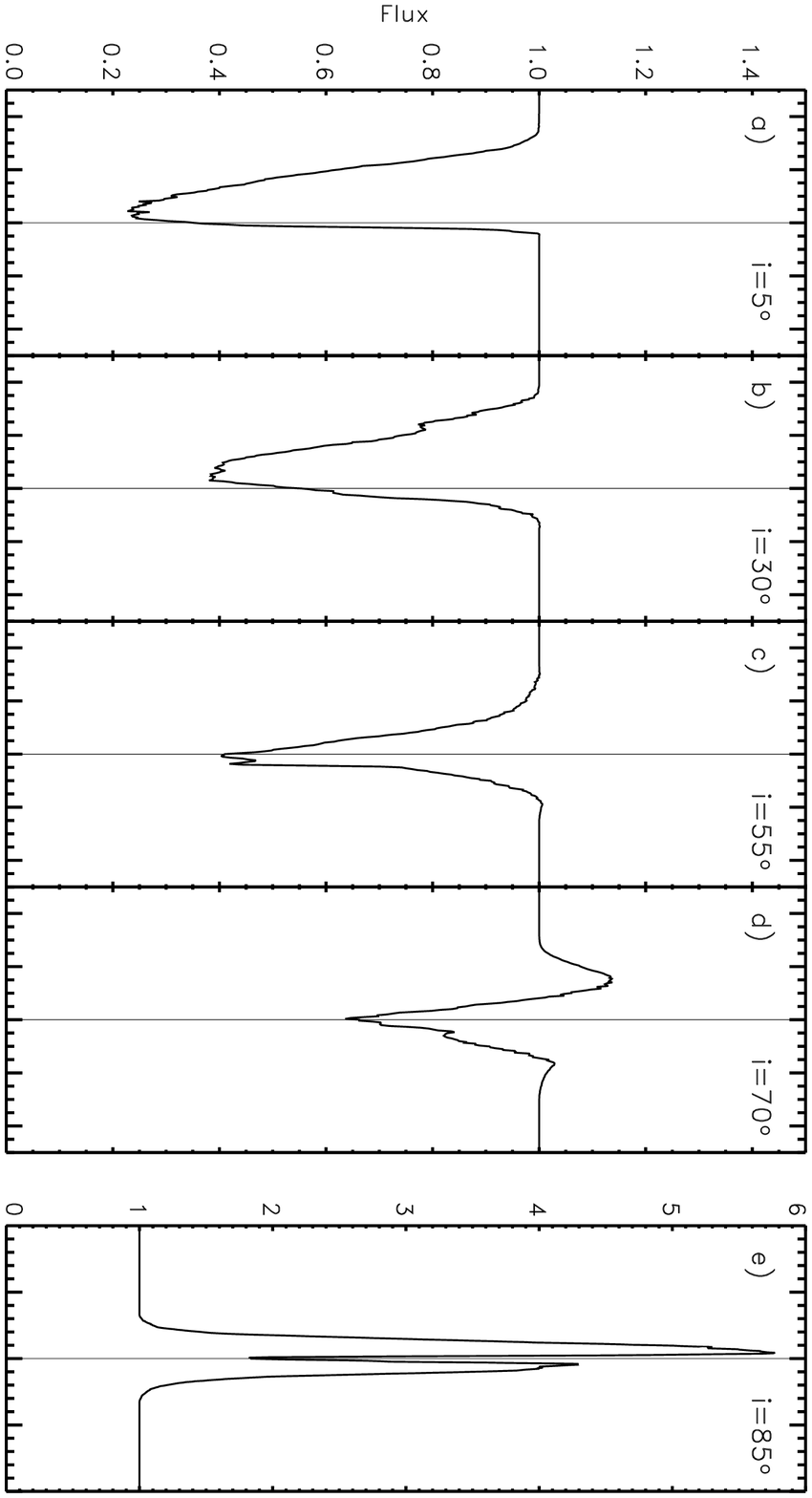}{0cm}{90}{26}{33}{190}{115}
\plotfiddle{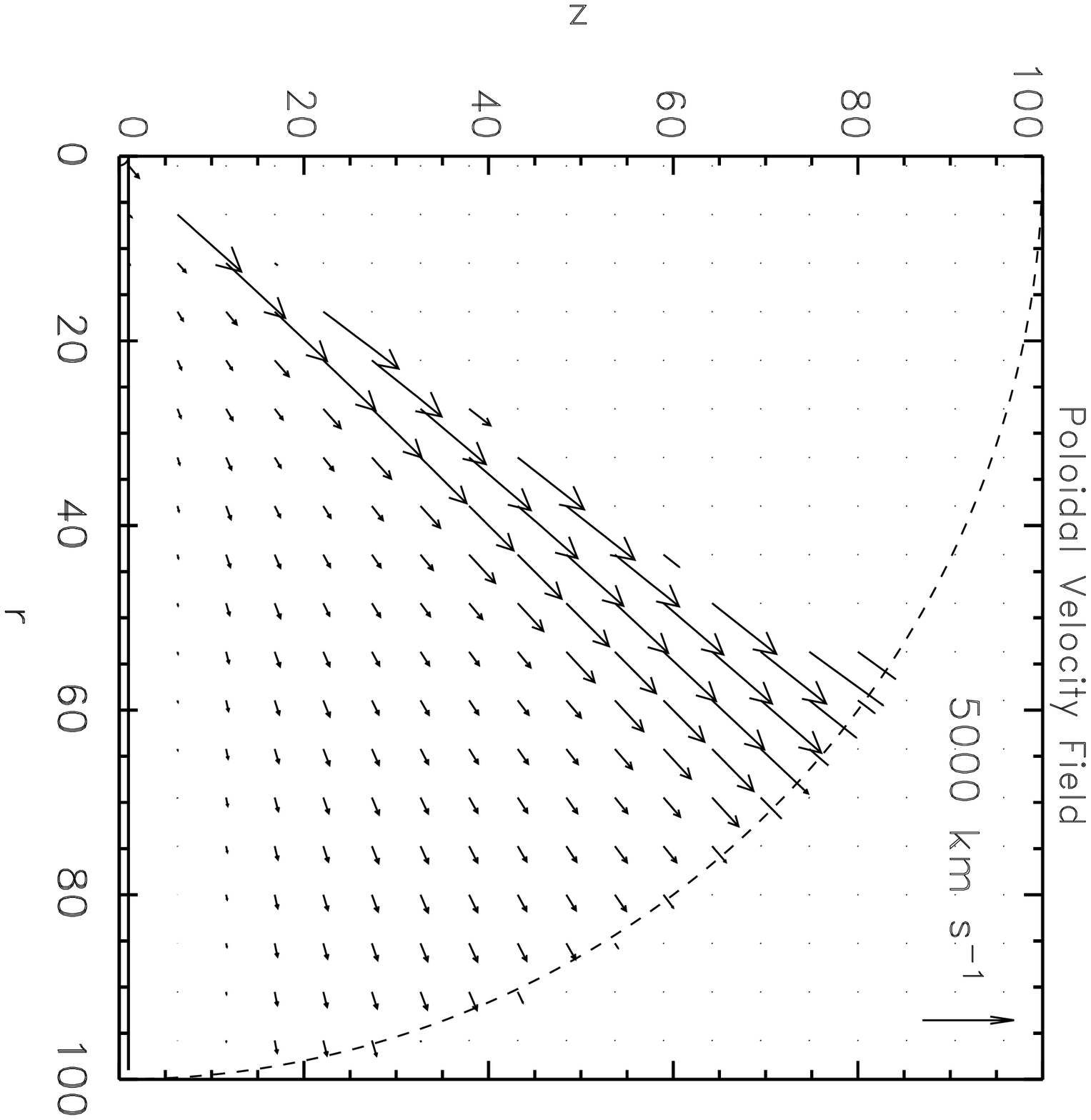}{0cm}{90}{28}{28}{30}{-20}
\plotfiddle{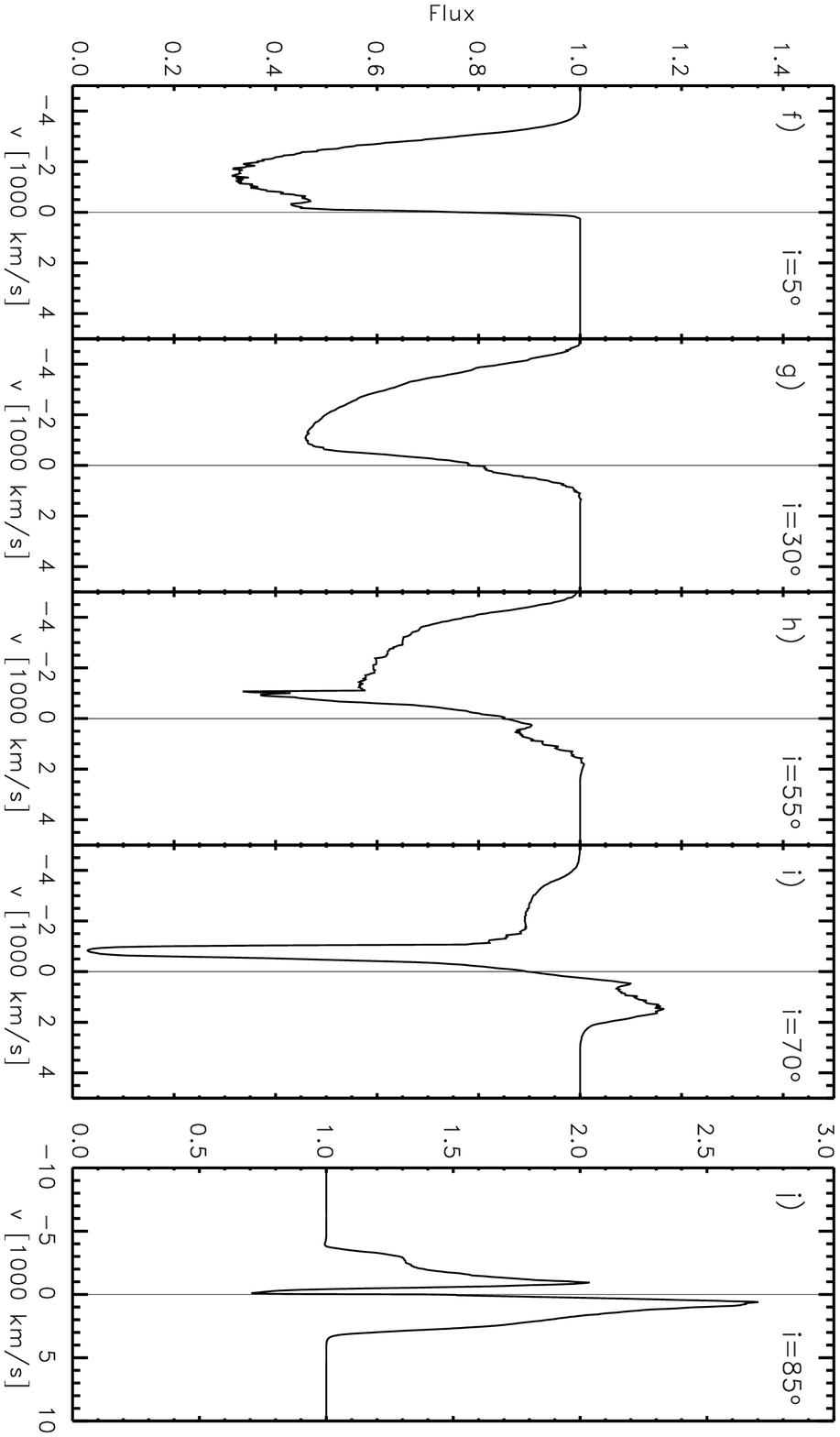}{0cm}{90}{26}{33}{190}{5} \caption{Maps
of poloidal velocity (the left hand side column of panels) and
line profiles (2-6 columns of panels) for two representative
models of LD disk winds. The line profiles are show for a range of
of inclination angle, $i$ (see top right corner of each panel for
the value of $i$). The top panels are for the wind model with
$x=0$ and $\dot{M}_{\rm a} = \pi \times 10^{-8}~\rm
M_{\odot}\,yr^{-1}$ [B2 model in \cite{P03b}]. The bottom panels
show results for the model with $\dot{M}_{\rm a} = \pi \times
10^{-8}~\rm M_{\odot}\,yr^{-1}$ and $x=1$ [C2 model in
\cite{P03b}]. The zero velocity corresponding to the line center
is indicated by the vertical line in the panels shown line
profiles. Note the difference in the velocity and flux ranges in
the planes for $i=85^\circ$ (the right hand side column).}
\end{figure}

Comparing the wind properties (e.g., $\dot{M}_{\rm w}$ and
geometry) inferred from observations with the wind properties
predicted by models for given systems parameters is one way of
testing the models. It would be very instructive to compute
synthetic spectra based on the models and compare these with
observed spectra. Such synthetic line profiles were computed by
\cite{Petal02}  using a generalized version of the Sobolev
approximation. In this study, the attention was restricted to the
case of a representative UV transition of a light ion such as C~IV
or Si~IV. The assumed abundance and atomic data were appropriate
to the C~IV$\lambda$1549 treated as singlet.

Generally, \cite{P02} found that the two main wind components
(slow wind and fast stream) produce distinct spectral features.
The fast stream produces profiles which show features consistent
with observations. These include the appearance of the classical
P-Cygni shape for a range of inclinations, the location of the
maximum depth of the absorption component at velocities less than
the terminal velocity, and the transition from net absorption to
net emission with increasing inclination. However the model
profiles have too little absorption or emission equivalent width
compared to  observed profiles. This quantitative difference
between the models and observations is not a surprise because, as
we discussed above, the LD wind models predict a mass loss rate,
mostly due to the fast stream, that is lower than the rate
required by the observations.

A key parameter shaping the total line profile -- made up of both
scattered emission and absorption -- is the ratio of the expansion
velocity to the rotational velocity.  This ratio is a cause of
some differences in the line profile between models with or
without radiation from the central star. For models with the WD
radiation switched on (the bottom panels on Figure 3), the winds
are less bipolar and so the rotational velocity decreases along
the wind streamlines faster than for models with the WD radiation
switched off (the top panels in Figure~3). This simple change in
the wind geometry reduces the rotational velocity of the flow
compared to the expansion velocity. The relatively higher
expansion velocity has an important consequence on the scattered
emission: for $x\simgreat 1$ at high inclination  ($i\simgreat
60^o$), the red component of the scattered emission becomes
stronger -- stronger than the blue component of the emission and
stronger than the blueshifted absorption so it is strong enough
for  the total line to have a  P~Cygni profile (e.g., Figure 3i)

In the follow up paper, we found that  the inclusion of a disk
wind at larger radii changes qualitatively and quantitatively the
line profiles predicted by the  LD disk wind model \citep{P03b}.
The models computed on a small grid -- such as  those in
\citeauthor*{PSD99}, where the outer radius equals 10 WD radii --
suffice to calculate the gross properties of the disk wind.
For example, the radial range of 10 WD radii suffice to
calculate $\dot{M}_{\rm w}$ and the fast part of the  wind,
which are both associated with the outflow from the innermost
disk. However, such calculations do not capture the entire region
where lines are formed. As a result, they underpredict the line
absorption and to a lesser extent the scattered emission. The
simulations on the small grid predicted a double-humped structure
near the line center for intermediate inclinations
\citep[e.g.,][]{P02}. This structure is due to a non-negligible
red-shifted absorption that is formed in the slow  wind where the
rotational velocity dominates over expansion velocity.

In \cite{P03b}, we showed  that by taking into account the
downstream part of the same slow  wind one is able to increase
significantly the central absorption. As a result, the
double-humped structure is reshaped to a more typical broad
trough. We emphasize that all improvements in the shape as well as
the strength of the absorption were achieved without changing the
gross properties of the wind. In particular, our new models do not
predict a higher mass-loss rate than the previous models. The
changes in the line profiles are mainly caused by the fact that
the ratio between the rotational and poloidal velocity decreases
downstream. Overall, one finds that the wind-formed line profiles
seen at ultraviolet wavelengths cannot originate in a flow where
rotation and poloidal expansion are comparable. The UV lines must
trace gas that expands substantially faster than it rotates.

The main discrepancy between the predicted line profiles and the
observed ones is in the line emission. Specifically, the model
cannot produce the redshifted emission as strong as that seen, for
example, in the C{\sc iv} profile of many systems with
intermediate inclinations (see below). However, this shortcoming
is not a great surprise -- this has been a problem for a while
\citep[see e.g.,][and discussion below]{MLK,Ketal96}.

A systematic comparison between predicted line profiles and
observations for many systems is a crucial test the idea
that the LD disk wind model can work for CV winds in the sense
that it can reproduce the observed line profiles for model parameters
(e.g., $L$) suitable to CVs. Preliminary results from limited survey of
dynamical models and their predictions are promising.
For example, Figure~4 presents a comparison between profiles
derived from the model with $\dot{M}_{\rm a}=\pi\times10^{-8}~\rm
M_{\odot}\,yr^{-1}$, $x=0.25$ [model E2 in \cite{P03b} for which
$\dot{M}_{\rm w}=8\times10^{-8}~\rm M_{\odot}\,yr^{-1}$] and
observations of the C{\sc iv}~1549\AA\ transition of  IX~Vel
\citep{Hetal}. [To show how much line emission is required, only
the absorption component is plotted.] The observed $i$  for this
system is $60^\circ$ \citep{BT}.

\begin{figure}[!t]
\plotfiddle{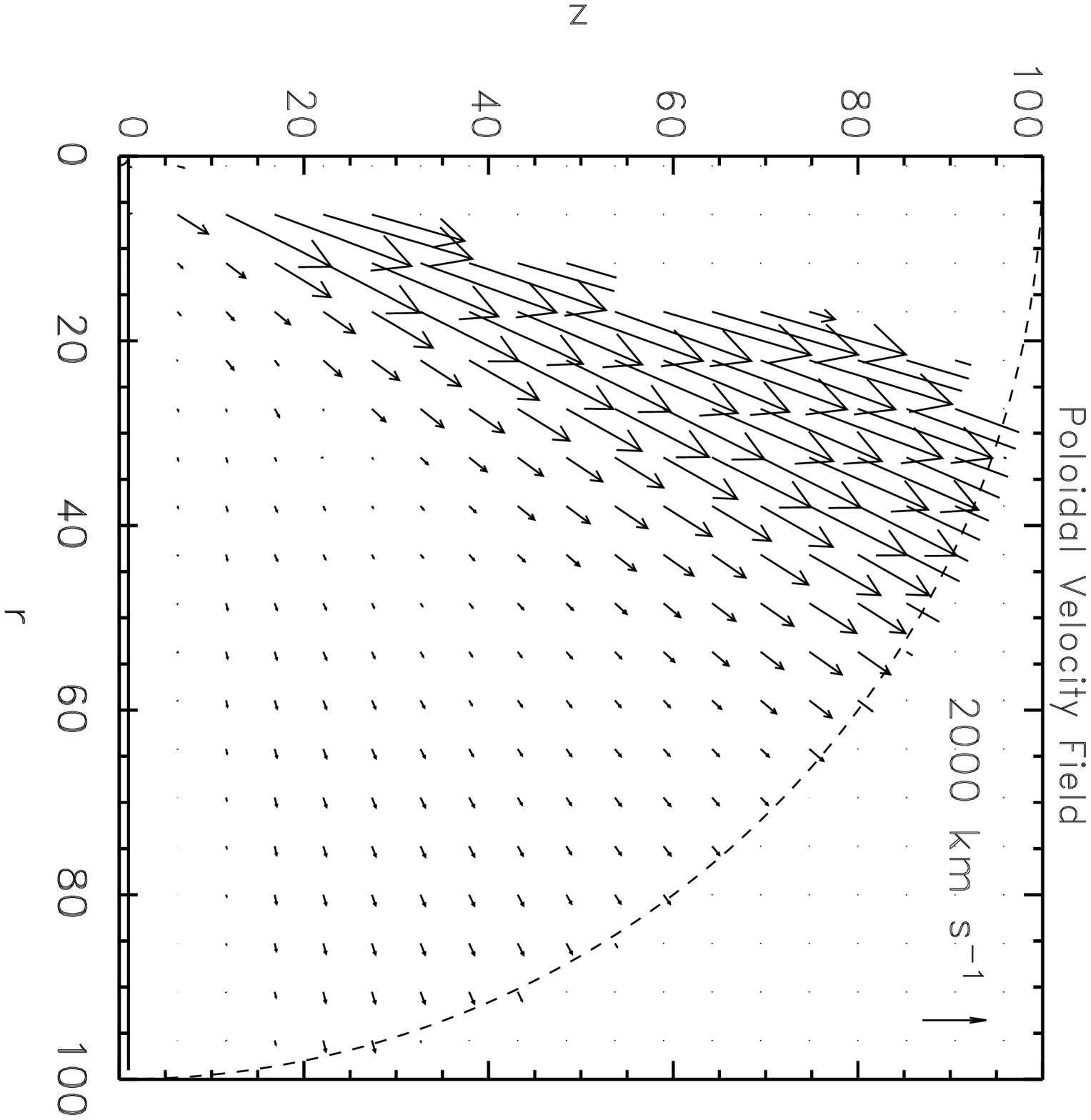}{4.0cm}{90}{29}{29}{40}{-30}
\plotfiddle{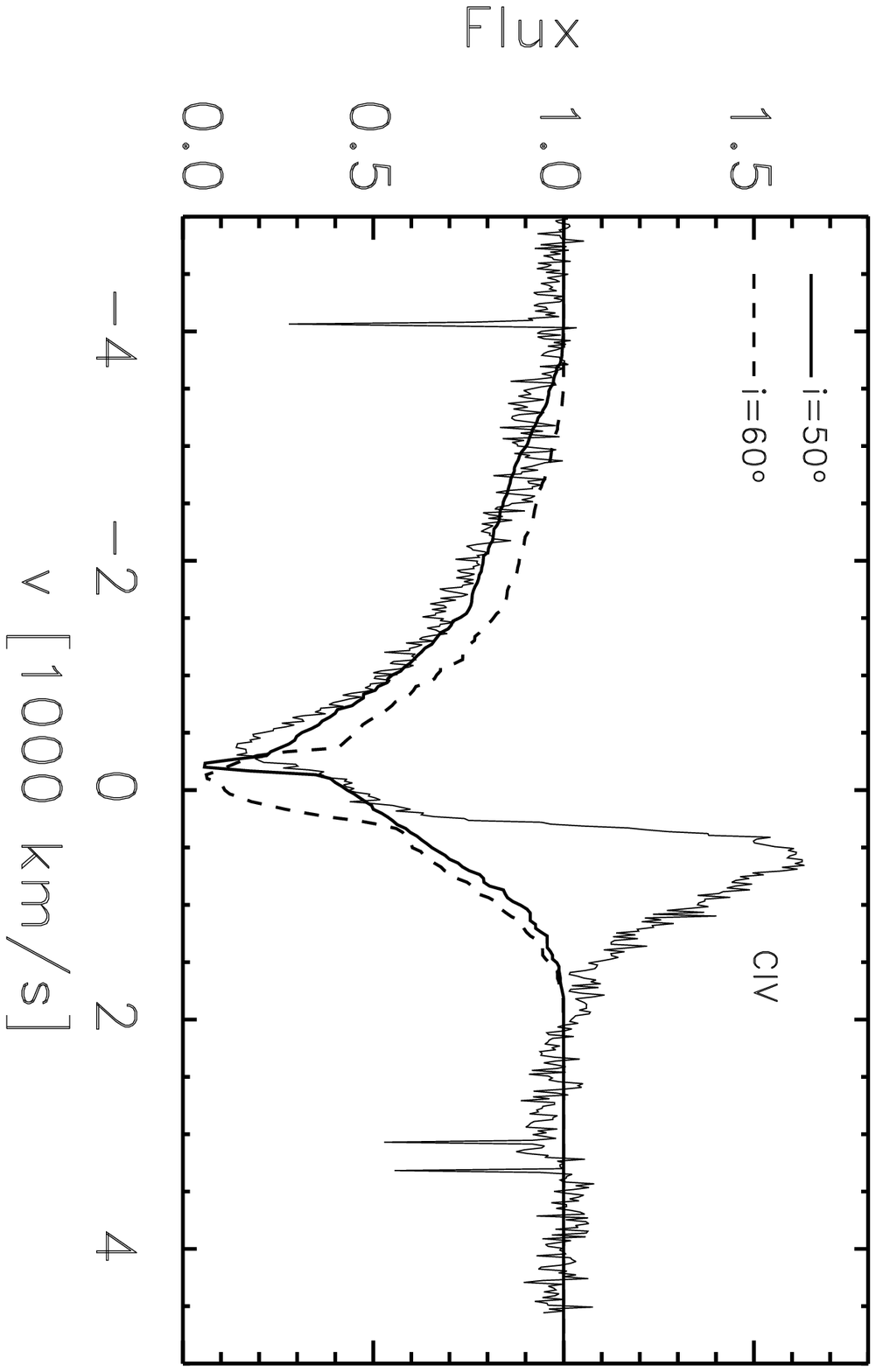}{0cm}{90}{35}{40}{230}{-20} \caption{A
map of poloidal velocity and line profiles. The right hand side
panel, compares profiles derived from a wind model for
$i=50^\circ$ and $60^\circ$ (thick solid and thick dashed line,
respectively) and observations of the C{\sc iv}~1549\AA\ in the
spectrum of the brightest nova-like variable, IX~Vel
\citep{Hetal}.  The synthesized lines show the absorption
component without the contribution from the scattered emission
(see the main text). }
\end{figure}

Figure 4 shows that the model profiles well reproduce the
blue-shifted absorption despite  a relatively low $\dot{M}_{\rm
w}$. Thus the gap between the kinematics of the LD wind models and
reality is narrower than comparison between observed and
theoretical $\dot{M}_{\rm w}$'s would indicate. The gap is only
narrowed but not bridged yet because the mass fluxes required to
match the observed spectra at least of IX~Vel are somewhat higher
than those observed. As \cite{DP00} discussed, the luminosity of
the system requires a mass accretion rate of at most $10^{-8}~\rm
M_{\odot}\,yr^{-1}$, whereas the line profiles require
$\dot{M}_{\rm a}=\pi\times10^{-8}~\rm M_{\odot}\,yr^{-1}$ and
$x=0.25$, yielding a system luminosity higher than the observed
one by a factor of $\sim 4$. However, the main point here is that
this discrepancy is much smaller than it used to be (i.e., it was
more than 1 order of magnitude) and there is a good chance that it
can be reduced still further.

For example, in \cite{P03b} we have computed many models, changing
various model parameters such as $\dot{M}_{\rm a}$ and the
parameters of the force multiplier, $\alpha$ and $M_{\rm max}$. In
general, in \cite{P03b} found that there is a degeneracy in the
model parameters as far as line profiles are concerned
(\citeauthor*{PSD98} found an analogous degeneracy for the wind
properties). For example, models with slightly different
parameters -- such as $\dot{M}_{\rm a}$ and $\alpha$ -- produce
similar line profiles for different $i$. Additionally, a model
with with $\dot{M}_{\rm a}=10^{-8}~\rm M_{\odot}\,yr^{-1}$  and
$\alpha=0.674$ predicts very similar line profiles to model with
$\dot{M}_{\rm a}=\pi \times 10^{-8}~\rm M_{\odot}\,yr^{-1}$  and
$\alpha=0.6$ for the same $i$. Finally, we find that the product
$(1+x) L_{\rm D} M_{\rm max}$, not its individual factors, appears
to be a fundamental parameter determining the line profiles (i.e.,
their width and depth) for the parameter range applicable to CVs.
This has an important implication for LD models: to obtain a
theoretical fit as good as shown in Figure~4 for a fixed $i$, one
needs $(1+x) L_{\rm D} M_{\rm max} \sim 1.3 \times 10^5~\rm
L_\odot$ rather than specifically $x=0.25$, $M_{\rm max}=4400$,
and $L_{\rm D}=23.4~\rm L_\odot$ as for the model shown in
Figure~4.

The above mentioned successes of LD disk wind models are
encouraging but there are also problems. For example, if radiation
pressure powers the mass loss from CVs, the wind mass-loss rate
should increase with increasing system luminosity (e.g.,
\citeauthor*{PSD98}; \citeauthor*{P99}\citeyear{P99}).). However
recent HST observations of IX~Vel  and V3885~Sgr showed that the
wind spectral features do not correlate with the system luminosity
\citep{Hetal}. If confirmed, these observations seriously
challenge the pure LD disk wind scenario.

\subsection{Limitations of LD models and future work}

We conclude this section with a reminder of the limitations of the
present LD disk wind models. Many of the details of the models
depend on details of the assumptions about the disk and
microphysics of the wind. For example, the assumed form for the
background radiation field inevitably plays a role in the
comparison of observed profiles with synthetic profiles based on
any model, kinematic or dynamical.  However for line fitting based
on LD wind models, the adopted disk and WD radiation fields are
even more important because they determine all wind properties
except the initial Keplerian component of motion. Current models
adopt the dependence of the disk radiation on radius according to
the standard steady state disk model \citep[e.g.,][]{PR}. This
assumption is a good starting point but it should be remembered
that actual disks may only be very crudely  described by simple
theory or they may yield unexpected features such as chromospheric
emission.  In fact, spectral synthesis models of accretion disk
photospheres \citep[e.g.,][]{LH,WH,WO} have typically failed to
adequately reproduce observed energy distributions where direct
comparisons have been made \citep[see e.g.,][]{Letal94}.  There
has been a similar lack of success in past calculations of
accretion disk line emission.  Specifically, the accounting for
both the strengths of and the flux ratios among various emission
lines in a large ensemble of observed CVs has not been compelling
\citep{MLK,Ketal96}. These models either miss a crucial physical
component or employ an inappropriate physical assumption which
affects the predicted line emission. It is for this reason that,
in comparing model wind profiles with observation, we concern
ourselves more with the absorption component than with the
emission.

In line profile calculations,  \cite{P02} and \cite{P03b} assumed
that the ionization fraction of scattering species is constant
($\xi_{\rm ion}$ is set to unity). A proper allowance for a
variation of $\xi_{\rm ion}$ with position to values less than one
can only serve to weaken the overall line profile.  [Examples of
plots of this potentially very marked positional dependence may be
found in the work of \cite{SV} and \cite{LK}]. Therefore, it is
important to carry out time-dependent calculations of the wind
photionization structure. It is also important to perform
simulations in fully three dimensions to explore nonaxisymmetric
effects. The two photoionization and three dimensional effects are
most likely coupled (e.g., by the wind density) and it would be
essential to study them self-consistently.

\section{MHD Models}

One of the reasons for considering magnetic fields as an
explanation for winds from accretion disks is the fact
that magnetic fields are very likely crucial for the existence of
all accretion disks. The magnetorotational instability (MRI)
has been shown to be  a very robust and universal
mechanism to produce turbulence and the transport of angular
momentum in disks at all radii \citep{BH91, BH98}. It is therefore
likely that magnetic fields control mass accretion inside the disk
and play a  role in producing a disk wind.

In fact, magnetically driven winds from disks are the favored
explanation for the outflows in many astrophysical environments.
\cite{BP} \citep[see also ][]{PP} showed that  the centrifugal
force can drive a wind from  the disk if the poloidal component of
the magnetic field, ${\bf B_{\rm p}}$ makes an angle of $> 30^o$
with respect to the normal to the disk surface. Generally,
centrifugally-driven MHD  disk winds (magnetocentrifugal winds for
short) require  the presence of a sufficiently strong,
large-scale, ordered magnetic field threading the disk with a
poloidal component at least comparable to the toroidal magnetic
field, $|B_\phi/B_{\rm p}| \simless 1$ \citep[e.g.,][]{CP,PP}.
Several groups have studied numerically axisymmetric winds
using the Blandford \& Payne mechanism
\citep[e.g.,][]{Uetal,OP97a,OP97b,KLB,KKS}. An important feature
of  magnetocentrifugal winds is that they require some assistance
to flow freely and steadily from the surface of the disk, to pass
through a slow magnetosonic surface \citep[e.g.,][]{BP}. The
numerical studies mentioned above do not resolve the vertical
structure of the disk but treat it as a boundary surface through
which mass is loaded on to the magnetic field lines at a specified
rate.

Winds from disks can also  driven by the magnetic pressure. In
particular, the toroidal magnetic field can quickly builds up due
to the differential rotation of the disk so that $|B_\phi/B_{\rm
p}| \gg 1$. In such a case, the magnetic pressure of the toroidal
field can give rise to a self-starting wind
\citep[e.g.,][]{US,PN,SN94,C95,KS}. To produce a steady outflow
driven by the magnetic pressure a steady supply of advected
toroidal magnetic flux at the wind base is needed, otherwise the
outflow is likely to be transient \citep[e.g.,][]{K93,C95}. It is
still not clear whether the differential rotation of the disk can
produce such a supply of the toroidal magnetic flux to match the
escape of magnetic flux in the wind and even if it does whether
such a system will be stable \citep[e.g.,][and references
therein]{C95,OP97b}.

To our best knowledge, in \cite{P03a}, we were first to report on
numerical simulations of the two-dimensional, time-dependent MHD
structure of  LD winds  from luminous accretion disks initially
threaded by a purely axial magnetic field. We developed
self-consistent models of such winds and applied them to  winds
from CVs. Our models require less free parameters than previous
MHD models. In particular, the model predicts the mass loss rate.

In \cite{P03a}, we used ideal MHD to compute the evolution of
Keplerian disks, varying the magnetic field strength and $L_{\rm
D}$, $L_\ast$, or both. We found  that the magnetic field very
quickly starts deviating from purely axial due to MRI. This leads
to fast growth of the toroidal magnetic field as field lines wind
up due to the disk rotation. As a result the toroidal field
dominates over the poloidal field above the disk and the gradient
of the former drives a slow and dense disk outflow, which
conserves specific angular momentum of fluid.

Our LD-MHD simulations also showed that depending on the strength
of the magnetic field relative to $L$ the disk
wind can be LD or MHD-driven. For very weak magnetic fields,
similarity to the LD wind, the wind consists of a dense, slow
outflow that is bounded on the polar side by a high-velocity
stream. The mass-loss rate is mostly due to the fast stream. As
the magnetic field strength increases first the slow part of the
flow is affected, namely it becomes even denser and slightly
faster and begins to dominate the mass-loss rate. In very strong
magnetic field or pure MHD cases, the wind consists of only a
dense, slow outflow without the presence of the distinctive fast
stream so typical of pure LD winds. Our simulations indicate that
winds launched by the magnetic fields are likely to remain
dominated by the fields downstream because of their relatively
high densities. Line driving may not be able to change a dense MHD
wind because the line force strongly decreases with increasing
density.

The increase of the mass loss rate due to the MHD effects is a
welcome development in modeling CV winds. However, it is  unclear
whether LD-MHD models can resolve the problem of too low
$\dot{M}_{\rm w}$. As we discussed in section~2, to explain CV
winds we need a model that predicts not only a higher
$\dot{M}_{\rm w}$ but also $\dot{M}_{\rm w}$ must be mostly due to
a fast wind not a dense slow rotating wind as we found in our
LD-MHD models.

\subsection{Limitations of LD-MHD models and future work}

The most important limitation of LD-MHD simulations is an
inadequate spatial resolution for modeling the MRI inside the
disk. These are only preliminary simulations we aimed to examine
the parameter space of the models that will define the major
trends in disk wind behavior. Therefore the priority has been so
far to set up the simulations in such a way that the base of the
wind is relatively stable and corresponds to a steady state
accretion disk. 

The fact that LD-MHD results strongly depend on the magnetic field
points to a need to explore different configurations for the
initial magnetic field and to move from two-dimensional
axisymmetric simulations to fully three-dimensional simulations.
It is important to follow the long-time evolution of the flow.
Therefore three-dimensional simulations are required as there
exist no self-sustained axisymmetric dynamos. Thus, contrary to
the stellar winds, simulations of winds from magnetized
disks -- with or without radiation pressure -- should include the
disks themselves, not just the disk photosphere, and should be
performed in three dimensions.

\section{Conclusion}
Our main conclusion is that, despite some problems, line-driving
alone is still the most plausible mechanism for driving the CV
winds. Preliminary results from LD-MHD wind models confirm that
magnetic driving is likely an important element of the wind
dynamics. However, magnetic driving does not seem to be necessary
to produce a wind. The most important issues which need to be
addressed by future dynamical models, regardless of driving
mechanism, are the effects of the position-dependent
photoionization and the dynamical effects in three dimensions.

\acknowledgments{We acknowledge support from NASA under LTSA grant
NAG5-11736. We also acknowledge support provided by NASA through
grant  AR-09947 from the Space Telescope Science Institute, which
is operated by the Association of Universities for Research in
Astronomy, Inc., under NASA contract NAS5-26555.}

\end{document}